\documentstyle[12pt,epsfig]{article}
\begin{document}
%
%\begin{titlepage}
%\pagestyle{empty}
%\vspace*{2cm}
\begin{center}
{\large\bf Spin Dependent Structure Function $g_1$ at Low $x$}\footnote{Talk 
given at the  Cracow Epiphany Conference on Spin Effects in Particle 
Physics, 9-11 January 1998, Cracow, Poland.}   
\vspace{1.1cm}\\ 
         {\sc J.~Kwieci\'nski}\footnote{e-mail: jkwiecin@solaris.ifj.edu.pl}\\ 
\vspace{0.3cm}
{\it Department of Theoretical Physics, \\
H.~Niewodnicza\'nski Institute of Nuclear Physics, 
Cracow, Poland}
\end{center}
\vspace{1.1cm}
\begin{abstract}  
 Theoretical expectations concerning  small $x$ behaviour of the spin 
 dependent 
structure function  $g_1$ 
are  summarised.  This includes discussion of  the Regge pole model 
predictions and of the double $ln^2(1/x)$ effects implied by perturbative 
QCD. The  quantitative implementation of the latter is described within 
the  unified scheme 
incorporating both Altarelli-Parisi evolution and the double $ln^2(1/x)$ 
resummation. The double $ln^2(1/x)$ effects are found to be  important in the 
region 
of $x$ which can possibly be probed at HERA.  Predictions for the polarized 
gluon distribution $\Delta G(x,Q^2)$ at low $x$ are also given.   
\end{abstract}
%\end{titlepage}
\vspace{1.1cm}
Understanding of the small $x$ behaviour of the spin dependent structure functions 
of the nucleon, where $x$ is the Bjorken parameter is interesting 
both theoretically and phenomenologically.  
Present experimental measurements  do not cover the low $x$ values and so 
the knowledge 
of  reliable extrapolation of the structure functions 
into this region is important for  estimate of  integrals which appear 
in the Bjorken and Ellis-Jaffe sum rules \cite{ALTAR}.  Theoretical description 
of the structure function $g_1^p(x,Q^2)$ at low $x$ is also extremely 
relevant for the possible polarised HERA measurements \cite{ALBERT}. 
The purpose of this talk is to  summarize the theoretical QCD 
expectations 
concerning the small $x$ behaviour of the spin dependent structure function 
$g_1$ of the nucleon. After brief reminder of the Regge pole expectations 
we shall discuss the effects of the double $ln^2(1/x)$ resummation 
and its phenomenological implementation. Besides the structure function $g_1$ 
we shall also discuss the spin dependent gluon distribution $\Delta G$. \\

The small $x$ behaviour of spin dependent structure functions  
for fixed $Q^2$ reflects the high energy behaviour of 
the 
virtual Compton scattering (spin dependent) total cross-section with 
increasing total CM energy 
squared $W^2$ since $W^2=Q^2(1/x-1)$. This is, by definition, the Regge limit 
and so the Regge pole exchange picture    
 \cite{PC} is 
therefore quite appropriate for the theoretical description 
of this behaviour. Here as usual $Q^2=-q^2$ where $q$ is the four momentum transfer 
between 
the scattered leptons. The relevant Reggeons which describe the small 
$x$ behaviour of the spin dependent structure functions 
are those which correspond to the axial vector mesons \cite{IOFFE,KARL}. 

The Regge pole model gives the
following small $x$ behaviour of the structure functions
$g_1^{i}(x,Q^2)$  where $g_1^{i}(x,Q^2), i=s,ns$ 
denote either  singlet  ($g_1^s(x,Q^2)=
g_1^p(x,Q^2)+g_1^n(x,Q^2)$)  or non-singlet ($g_1^{ns}(x,Q^2)=
g_1^p(x,Q^2)-g_1^n(x,Q^2)$) combination of structure functions:

\begin{equation}
g_1^{i}(x,Q^2) = \gamma_i(Q^2)x^{-\alpha_{i}(0)}
\label{rg1}
\end{equation}
where $g_1^{i}(x,Q^2)$ 
denote either  singlet  ($g_1^s(x,Q^2)=
g_1^p(x,Q^2)+g_1^n(x,Q^2)$)  or non-singlet ($g_1^{ns}(x,Q^2)=
g_1^p(x,Q^2)-g_1^n(x,Q^2)$) combination of structure functions.  

\noindent 
In eq. and (\ref{rg1})  
$\alpha_{s,ns}(0)$ denote the intercept of the Regge pole
trajectory corresponding to the axial vector mesons with $I=0$
or $I=1$ respectively. It is expected that   $\alpha_{s,ns}(0) \le
0$ and that $\alpha_{s}(0) \approx \alpha_{ns}(0)$ i.e. the singlet 
spin dependent structure function is expected to have similar low $x$ 
behaviour as the non-singlet one.\\

It may be instructive to confront this behaviour with the Regge pole 
expectations for the spin independent structure function $F_1(x,Q^2)$:  
   
\begin{equation}
F_1^i(x,Q^2)= \beta_i(Q^2) x^{-\alpha_i(0)}. 
\label{reggef}
\end{equation}
 The singlet part $F_1^{S}=F_1^p+F_1^n$
of the structure function $F_1$ is controlled at small $x$ by 
pomeron exchange, while the non-singlet part $F_1^{NS}=
F_1^p-F_1^n$ by  $A_2$ reggeon.  The pomeron intercept is significantly 
different from that of the $A_2$ Reggeon i.e.  $\alpha_P(0)=1+\epsilon$  
with $\epsilon \approx 0.08$ while $\alpha_{A_2}(0) \approx 0.5$ as determined 
from the fits to the total hadronic and photoproduction data \cite{DOLA}.  
 This implies 
that in the spin independent case  the singlet part $F_1^S(x,Q^2)$ of the structure 
function $F_1(x,Q^2)$ dominates  at low $x$ over the non-singlet component.\\

Several  of the Regge pole model expectations 
for both spin dependent and spin independent structure functions  
are modified by  perturbative QCD effects.  In particular as far as the 
spin dependent structure functions are concernced the Regge behaviour 
(\ref{rg1})  becomes unstable against the QCD evolution which generates 
more singular behaviour than that given by eq. (\ref{rg1}) for $\alpha_{i}(0)
 \le 0$.  In the LO approximation one gets: 
$$
g_1^{NS}(x,Q^2) \sim exp [2 \sqrt{\Delta P_{qq}(0) \xi(Q^2) ln(1/x)}]
$$

\begin{equation}
g_1^S(x,Q^2) \sim  exp [2 \sqrt{\gamma^{+} \xi(Q^2) ln(1/x)}] 
\label{dlogq2}
\end{equation}
where 
\begin{equation}
 \xi(Q^2) = \int_{Q_0^2}^{Q^2}{dq^2\over q^2} {\alpha_s(q^2)\over 2 \pi} 
\label{xi}
\end{equation}
and 
\begin{equation}
\gamma^+ = {\Delta P_{qq}(0) +\Delta P_{gg}(0) + 
\sqrt{(\Delta P_{qq}(0) -\Delta P_{gg}(0))^2 + 
4\Delta P_{qg}(0)\Delta P_{gq}(0)}\over 2}
\label{lplus}
\end{equation}
with $\Delta P_{ij}(0) = \Delta P_{ij}(z=0)$ where $\Delta P_{ij}(z)$ 
denote the LO splitting functions describing evolution 
of spin dependent  parton densities. To be precise we have: 
\begin{eqnarray}
\Delta P_{qq}(0) &=& {4\over 3} 
\\ \nonumber
\Delta P_{qg}(0) &=& - N_F 
\\ \nonumber
\Delta P_{gq}(0) &=& {8\over 3} 
\\ \nonumber
\Delta P_{gg}(0) &=& 12.  
\label{dpij}
\end{eqnarray}

We recall for comparison the analogous small $x$ behaviour of the 
structure function $F_1(x,Q^2)$ in the LO approximation : 
\begin{equation}
F_1^S(x,Q^2) \sim {1\over x}exp [2 \sqrt{6 \xi(Q^2) ln(1/x)}]. 
\label{dlqf1}
\end{equation}
The Regge behaviour of the non-singlet structure function $F_1^{NS}(x,Q^2)$ 
remains stable 
against the QCD evolution.\\

The LO (and NLO) QCD evolution which sums the leading (and next-to-leading) 
powers of $lnQ^2/Q_0^2)$ is however incomplete at low $x$.  In this region  
 one should worry about another "large" logarithm which is $ln(1/x)$ and 
 resum its leading powers.  In the spin independent case this is provided 
 by the Balitzkij, Fadin, Kuraev, Lipatov (BFKL) equation \cite{BFKL} which 
 gives in the leading $ln(1/x)$ approximation the following small $x$ behaviour 
 of $F_1^S(x,Q^2)$ 
\begin{equation}
F_1^S(x,Q^2) \sim x^{-\lambda_{BFKL}} 
\label{bfkl1}
\end{equation}
where the intercept of the BFKL pomeron $\lambda_{BFKL}$ is given in the leading order by the 
following formula:
\begin{equation} 
\lambda_{BFKL}=1+{3\alpha_s\over \pi}4ln(2). 
\label{bfkl2}
\end{equation}

It has recently been pointed out that the spin dependent structure function 
$g_1$ at low $x$ is dominated  by the {\it double} logarithmic $ln^2(1/x)$ 
contributions 
i.e. by those terms of the perturbative expansion which correspond to 
the powers of $ln^2(1/x)$ at each order of the expansion 
\cite{BARTNS,BARTS}.  Those contributions go {\it beyond} 
the LO or NLO order QCD evolution of polarised parton densities \cite{AP} and in order to take 
them into account one has to 
include the resummed double $ln^2(1/x)$ terms in the coefficient 
and splitting functions \cite{RG}. In this talk we will present discussion 
of the double $ln^2(1/x)$ resuumation following alternative approach based  
on unitegrated distributions \cite{BBJK,BZIAJA}.\\

The dominant contribution to the double $ln^2(1/x)$ resummation comes from 
the ladder diagrams with quark and gluon exchanges along the ladder (cf. Fig.1).  
In what follows we shall neglect for simplicity 
possible 
non-ladder bremsstrahlung terms which are relatively unimportant 
\cite{BARTNS,BARTS}.\\

%FIGURE 1
\begin{figure}[htb]
   \vspace*{-1cm}
    \centerline{
     \psfig{figure=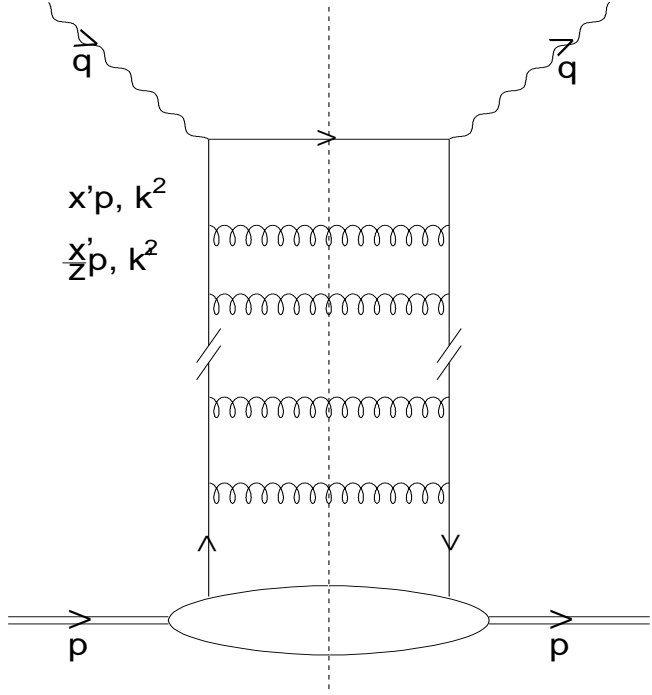,height=12cm,width=14cm}
               }
    \vspace*{-0.5cm}
     \caption{An example of a ladder diagram generating 
     double logarithmic terms in the  spin structure function $g_1$.}
\label{fig.1}
\end{figure} 
It is convenient to introduce the unintegrated (spin dependent) parton 
distributions 
$f_i(x^{\prime},k^2)$ ($i=u_{v},d_{v},\bar u,\bar d,\bar s,g$) where $k^2$ 
is the transverse momentum squared of the parton $i$ and $x^{\prime}$ the 
longitudinal momentum fraction of the parent nucleon carried by a parton.  
The conventional (integrated) distributions $\Delta p_i(x,Q^2)$ are 
related in the 
following way  to the unintegrated distributions $f_i(x^{\prime},k^2)$: 
\begin{equation}
 \Delta p_i(x,Q^2)=\Delta p_i^{(0)}(x)+
 \int_{k_0^2}^{W^2}{dk^2\over k^2}f_i(x^{\prime}=
x(1+{k^2\over Q^2}),k^2)
\label{dpi}
\end{equation}
where $\Delta p_i^{(0)}(x)$ is the nonperturbative part of the 
of the distributions.  
The parameter $k_0^2$ is the infrared cut-off which will be set equal 
to 1 GeV$^2$. 
The origin of the nonperturbative part $\Delta p_i^{(0)}(x)$ can be viewed
upon as originating from the non-perturbative region  $k^2<
k_0^2$, i.e. 
\begin{equation}
\Delta p_i^{(0)}(x)= \int_{0}^{k_0^2}{dk^2\over k^2}f_i(x,k^2)    
\label{gint0}
\end{equation} 
The spin dependent structure function $g_1^p(x,Q^2)$ of the proton is 
related in a standard 
way to the (integrated) parton distributions:
$$
g_1^p(x,Q^2)=
$$
\begin{equation}
{1\over 2}\left[{4\over 9}(\Delta u_v(x,Q^2) + 2\Delta \bar u (x,Q^2))+
{1\over 9}(\Delta d_v(x,Q^2) + 2\Delta \bar u(x,Q^2) +  2\Delta \bar s(x,Q^2))
\right]
\label{gp1}
\end{equation}                           
where $\Delta u_v(x,Q^2) = \Delta p_{u_v}(x,Q^2)$ etc.  
We assume $\Delta \bar u =
\Delta \bar d$ and confine ourselves to the number of flavours $N_F$ equal to 
three.\\

The valence quarks distributions and asymmetric part of the sea 
\begin{equation} 
f_{us}(x^{\prime},k^2)= f_{\bar u}(x^{\prime},k^2)-f_{\bar s}(x^{\prime},k^2)
\label{us}
\end{equation}  
will correspond to  ladder diagrams with quark exchange along the ladder.  
The singlet distributions 
\begin{equation}
f_{S}(x^{\prime},k^2) = f_{u_v}(x^{\prime},k^2)+f_{d_v}(x^{\prime},k^2)+
4f_{\bar u}(x^{\prime},k^2)+ 
2f_{\bar s}(x^{\prime},k^2)
\label{sns}
\end{equation}
and the gluon distributions $f_g(x^{\prime},k^2)$ will correspond to ladder 
diagrams with both quark (antiquark) and gluon exchanges along the ladder. \\
   
The sum of double logarithmic $ln^2(1/x)$ 
terms corresponding to ladder diagrams is generated by the following 
integral equations (see Fig.1): 
\begin{equation} 
f_{k}(x^{\prime},k^2)=f^{(0)}_{k}(x^{\prime},k^2) + 
{\alpha_s\over 2 \pi} \Delta P_{qq}(0)   
\int_{x^{\prime}}^1 {dz\over z} 
\int_{k_0^2}^{k^2/z}
{dk^{\prime 2}\over k^{\prime 2}} 
f_{k}\left({x^{\prime}\over z},k^{\prime 2}\right)
\label{dlxns}
\end{equation}
($k= u_v, d_v, us$)

$$ 
f_{S}(x^{\prime},k^2)=f^{(0)}_{S}(x^{\prime},k^2) +
$$ 
$$ 
{\alpha_s\over 2 \pi}    
\int_{x^{\prime}}^1 {dz\over z} 
\int_{k_0^2}^{k^2/z}
{dk^{\prime 2}\over k^{\prime 2}} 
\left[\Delta P_{qq}(0)
f_{S}\left({x^{\prime}\over z},k^{\prime 2}\right)+
\Delta P_{qg}(0)
f_{g}\left({x^{\prime}\over z},k^{\prime 2}\right)\right]
$$

$$ 
f_{g}(x^{\prime},k^2)=f^{(0)}_{g}(x^{\prime},k^2) + 
$$
\begin{equation}
{\alpha_s\over 2 \pi}    
\int_{x^{\prime}}^1 {dz\over z} 
\int_{k_0^2}^{k^2/z}
{dk^{\prime 2}\over k^{\prime 2}} 
\left[\Delta P_{gq}(0)
f_{S}\left({x^{\prime}\over z},k^{\prime 2}\right)+
\Delta P_{gg}(0)
f_{g}\left({x^{\prime}\over z},k^{\prime 2}\right)\right]
\label{dlxsg}
\end{equation} 
with $\Delta P_{ij}(0) = \Delta P_{ij}(z=0)$ given by eq. (\ref{dpij}).  

The variables $k^2$($k^{\prime 2})$ denote the transverse momenta 
squared of the
quarks (gluons) exchanged along the ladder, $k_0^2$ is  the infrared cut-off and the inhomogeneous terms
$f_i^{(0)}(x^{\prime},k^2)$ will be specified later. 
The integration limit $k^2/z$ 
follows from the requirement that the  virtuality of the
quarks (gluons) exchanged along the ladder is controlled by 
the tranverse momentum squared.\\

The origin of the double logarithmic ${\rm ln}^2(1/x)$ terms in $g_1(x,Q^2)$  
can be traced to the fact that the conventional single logarithmic terms
coming from the logarithmic integration over the longitudinal
momentum fraction $z$ are enhanced by the logarithmic
integration over the transverse momentum up to the $z$ dependent
limit $k^2/z$ in equations (\ref{dlxns},\ref{dlxsg}) and up to the $x$ 
dependent 
limit $W^2=Q^2(1/x-1)$ in eq. (\ref{dpi}).\\
    
Equation (\ref{dlxns}) is similar to the corresponding equation in QED 
describing the double logarithmic resummation generated by ladder diagrams with 
fermion exchange \cite{QED}. The problem of double logarithmic 
asymptotics in QCD in the non-singlet channels was also discussed in ref  
\cite{QCD}.\\

Equations (\ref{dlxns}, \ref{dlxsg}) generate singular power behaviour of the 
spin dependent parton distributions and of the spin dependent 
structure functions $g_1$  at small  $x$ i.e.  
$$ 
g_1^{NS}(x,Q^2) \sim x^{-\lambda_{NS}} 
$$ 
$$
g_1^{S}(x,Q^2) \sim x^ {-\lambda_{S}}
$$
\begin{equation}
\Delta G(x,Q^2) \sim x^ {-\lambda_{S}}
\label{power}
\end{equation}
where $g_1^{NS}=g_1^p-g_1^n$ and $g_1^{S}=g_1^p+g_1^n$ respectively and 
$\Delta G$ is the spin dependent gluon distribution.  This behaviour reflects 
 similar small $x^{\prime}$ behaviour of the unintegrated distributions.  
Exponents $\lambda_{NS,S}$ are given by the following formulas:  
$$
\lambda_{NS} = 2 \sqrt{\left[{\alpha_s \over 2\pi} \Delta P_{qq}(0)\right]}
$$
\begin{equation}
\lambda_{S} = 2 \sqrt{\left[{\alpha_s \over 2\pi} \gamma^+\right]}
\label{lambda}
\end{equation}
where $\gamma^+$ is given by eq.(\ref{lplus}). The power-like behaviour 
(\ref{power}) with the exponents $\lambda_{NS,S}$ given 
by eq. (\ref{lambda}) remains 
the leading small $x$ behaviour of the structure functions provided that 
their non-perturbative parts are less singular.  This takes place 
if the latter are assumed to have the Regge pole like behaviour with the 
corresponding intercept(s) being near $0$. \\

In order to understand  origin of the power-like behaviour (\ref{power}) 
it is useful 
to go to the moment space and inspect the singularities of the 
moment functions $\bar f_i(\omega,k^2)$ and $\Delta \bar p_i(\omega,Q^2)$  
$$
\bar f_i(\omega,k^2) = \int_0^1 dx^{\prime} x^{\prime \omega - 1} 
f_i(x^{\prime},k^2)
$$
\begin{equation}
\Delta \bar p_i(\omega,Q^2)=\int_0^1 dx x^{ \omega - 1} \Delta
p_i(x,Q^2)
\label{moments}
\end{equation}
in the $\omega$ plane.
It follows from eq. (\ref{dpi}) that the moment functions 
$\Delta \bar p_i(\omega,Q^2)$ 
are related in the following way to $\bar f_i(\omega,k^2)$:
\begin{equation}
\Delta \bar p_i(\omega,Q^2) = \Delta \bar p_i^{(0)}(\omega) + 
\int_{k_0^2}^{\infty} {dk^2\over k^2} \left(1+{k^2\over Q^2}\right)^{-\omega}
\bar f_i(\omega,k^2)
\label{momint}
\end{equation}
where $\Delta \bar p_i^{(0)}(\omega)$ denote the moment functions of the 
nonperturbative distributions $\Delta p_i^{(0)}(x)$. 
Let us first consider the case $i=u_v,d_v,us$.  
Equation (\ref{dlxns}) implies the following equation 
for the moment functions $\bar f_i(\omega,k^2)$ 
\begin{equation}
\bar f_i(\omega,k^2)=\bar f_{i}^{(0)}(\omega,k^2)+ 
{\bar \alpha_s \over \omega} \left[\int_{k_0^2}^{k^2} 
{dk^{\prime 2}\over 
k^{\prime 2}}\bar f_i(\omega,k^{\prime 2})+ 
\int_{k^2}^{\infty} {dk^{\prime 2}\over 
k^{\prime 2}}\left({k^2 \over k^{\prime 2}}\right)^{\omega}
\bar f_i(\omega,k^{\prime 2})\right]
\label{dleqm}
\end{equation}
In this equation $\bar f_{i}^{(0)}(\omega,k^2)$ denote the moment functions 
of  $f_{i}^{(0)}(x^{\prime},k^2)$ and $\bar \alpha_s$ is defined 
by :
\begin{equation}
\bar \alpha_s= \Delta P_{qq}(0) { \alpha_s \over 2 \pi}
\label{alphabar}
\end{equation}
Equation (\ref{dleqm}) follows from (\ref{dlxns}) after taking  into 
account the following relation: 
\begin{equation}
\int_0^1{dz\over z}z^{\omega}\Theta \left({k^2\over k^{\prime 2}} - z
\right)=
{1\over \omega}\left[\Theta(k^2-k^{\prime 2})+
\left({k^2\over k^{\prime 2}}\right)^
{\omega}\Theta (k^{\prime 2}-k^2)\right]. 
\label{theta}
\end{equation}
For fixed coupling $\bar \alpha_s$  equation (\ref{dleqm})
 can be solved analytically.  
Assuming for simplicity that the inhomogeneous term is independent 
of $k^2$ (i.e. that $\bar f_{i}^{(0)}(\omega,k^2) = C_i(\omega)$ )
we get the following solution of  eq.(\ref{dleqm}): 
\begin{equation}
\bar f_i(\omega,k^2)=R_i(\bar \alpha_s, \omega) 
\left({k^2\over k_0^2}\right)^{\tilde \gamma_{NS}(\alpha_s,\omega)}
\label{solm}
\end{equation}
where 
\begin{equation}
\tilde \gamma_{NS}(\alpha_s, \omega) = {\omega - \sqrt{\omega^2 - 4 \bar 
 \alpha_s}\over 2}
\label{anomd}
\end{equation}
and
\begin{equation}
R_i(\bar \alpha_s,  \omega)=C_i(\omega) {\tilde \gamma_{NS} (\alpha_s, \omega) \omega
\over 
\bar \alpha_s}. 
\label{r}
\end{equation}
Equation (\ref{anomd}) defines the non-singlet anomalous dimension in which 
 the double logarithmic $ln^2(1/x)$ terms i.e. the powers of ${\alpha_s \over 
\omega^2}$ have been resummed to all orders.  It can be seen from (\ref{anomd}) 
that this anomalous dimension has a (square root) branch point singularity 
at $\omega=
\lambda_{NS}$ 
\begin{equation}
\lambda_{NS}= 2 \sqrt{\bar \alpha_s}. 
\label{barom}
\end{equation} 
This singularity will of course be also present in the moment functions $
\bar f_i(\omega,k^2)$  and $\Delta \bar p_i(\omega, Q^2)$. 
It generates singular power - like behaviour of the non-singlet structure 
function $g_1^{NS}(x,Q^2)$ (cf. eq.(\ref{power})). For 
$C_i(\omega)={\bar \alpha_s\over \omega} 
\Delta \bar  p_i^{(0)}(\omega)$ the moment functions 
$\Delta \bar p_i(\omega, Q^2)$ can be shown  to have a familiar RG form  
\begin{equation}
\Delta \bar p_i(\omega, Q^2)=\bar R_i(\omega,\alpha_s) 
\left({Q^2\over k_0^2}\right)^{\tilde \gamma_{NS}(\alpha_s, \omega)} + 
O({k_0^2\over Q^2})
\label{rgns}
\end{equation}
where
\begin{equation}
 \bar R_i(\omega,\alpha_s)=\Delta \bar  p_i^{(0)}(\omega){\Gamma(\tilde 
 \gamma_{NS} (\alpha_s, \omega)+1))
\Gamma(\omega-\tilde \gamma_{NS}(\alpha_s, \omega))\over \Gamma(\omega)}
\label{rgr}
\end{equation}
It may be seen from eq. (\ref{rgr}) that the singularity at $\omega=\lambda_{NS}$ which is present 
in the anomalous dimension $\tilde \gamma_{NS} (\alpha_s, \omega)$ does also 
appear in 
the functions $\bar R_i$ .  It is therefore present in the moment 
functions $\Delta \bar p_i(\omega, Q^2)$ for arbitrary values of the 
scale $Q^2$ including $Q^2=k_0^2$.  The situation in this case is 
 similar to that in the solution of the BFKL 
equation \cite{CIAFAL}.\\

The formula for the exponent $\lambda_S$ which controls the small $x$ behaviour 
of $g_1^S(x,Q^2)$ and of $\Delta G(x,Q^2)$ can be obtained by inspecting the 
singularities in the $\omega$ plane of the moment functions 
 $\bar f_S(\omega,k^2)$ 
and $\bar f_g(\omega,k^2)$ which are generated by the corresponding equations 
for those functions.  They can be obtained from eqs.(\ref{dlxsg}) and have the 
following form: 

$$ 
\bar f_{S}(\omega,k^2)=\bar f^{(0)}_{S}(\omega,k^2) +
{\alpha_s\over 2 \pi \omega}    
\int_{k_0^2}^{\infty}
{dk^{\prime 2}\over k^{\prime 2}}\left[\Theta(k^2-k^{\prime 2}) + 
\Theta(k^{\prime 2}-k^{2}) 
\left({k^{2}\over k^{\prime 2}}\right)^{\omega}\right]
$$ 

$$ 
\left[\Delta P_{qq}(0)
\bar f_{S}(\omega,k^{\prime 2})+
\Delta P_{qg}(0)
\bar f_{g}(\omega,k^{\prime 2})\right]
$$

$$ 
\bar f_{g}(\omega,k^2)=\bar f^{(0)}_{g}\omega, k^2) + 
{\alpha_s\over 2 \pi \omega}\int_{k_0^2}^{\infty}
{dk^{\prime 2}\over k^{\prime 2}} 
\left[\Theta(k^2-k^{\prime 2}) + 
\Theta(k^{\prime 2}-k^{2}) 
\left({k^{2}\over k^{\prime 2}}\right)^{\omega}\right]
$$

\begin{equation}
\left[\Delta P_{gq}(0)
\bar f_{S}(\omega,k^{\prime 2})+
\Delta P_{gg}(0)
\bar f_{g}(\omega,k^{\prime 2})\right]. 
\label{dlxsgm}
\end{equation}

For $k^2$ independent inhomogeneous terms $(\bar f_{S,g}^{(0)}(\omega,k^2)=
C_{S,g}(\omega))$  
 these equations have the following solution:  
\begin{equation}
\bar f_{S,g}(\omega,k^2) = R_{S,g}^{+}(\omega,\alpha_s) 
\left({k^2\over k_0^2}\right)^{\bar 
\gamma^{+}(\omega,\alpha_s)} + R_{S,g}^{-}(\omega,\alpha_s) 
\left({k^2\over k_0^2}\right)^{\bar 
\gamma^{-}(\omega,\alpha_s)}
\label{solsing}
\end{equation}
The (singlet) anomalous dimensions $\tilde  
\gamma^{\pm}(\omega,\alpha_s)$ are given by the following equations:

\begin{equation}
\tilde  
\gamma^{\pm}(\omega,\alpha_s)= {\omega-\sqrt{\omega^2-4{\alpha_s\over 2 \pi}
\gamma^{\pm}}\over 2}
\label{gammapm}
\end{equation}
where $\gamma^+$ is defined by equation (\ref{lplus}) while $\gamma^-$ 
is: 
\begin{equation}
\gamma^- = {\Delta P_{qq}(0) +\Delta P_{gg}(0) -
\sqrt{(\Delta P_{qq}(0) -\Delta P_{gg}(0))^2 + 
4\Delta P_{qg}(0)\Delta P_{gq}(0)}\over 2}
\label{lminus}
\end{equation}
The functions $R^{\pm}_{S,g}(\omega,\alpha_s)$ can be expressed in terms 
of $C_{S,g}(\omega)$ and the anomalous dimensions $\tilde  
\gamma^{\pm}(\omega,\alpha_s)$.  The moment functions 
$\Delta \bar p_{S,g}(\omega,Q^2)$ can be shown to have the RG form:
\begin{equation}
\Delta \bar p_{S,g}(\omega,Q^2) = \bar R_{S,g}^{+}(\omega,\alpha_s) 
\left({Q^2\over k_0^2}\right)^{\bar 
\gamma^{+}(\omega,\alpha_s)} + \bar R_{S,g}^{-}(\omega,\alpha_s) 
\left({Q^2\over k_0^2}\right)^{\bar 
\gamma^{-}(\omega,\alpha_s)} + O({k_0^2\over Q^2})
\label{rgsg}
\end{equation}
It may be seen from eq.(\ref{gammapm}) that the anomalous dimensions       
$\bar \gamma^{\pm}(\omega,\alpha_s)$ have (branch point) singularities 
at $\bar \omega ^{\pm}$  
\begin{equation}
\bar \omega ^{\pm} = 2\sqrt{{\alpha_s\over 2 \pi}
\gamma^{\pm}} 
\label{ompm}
\end{equation}
with $\bar \omega^+ > \bar \omega^{-}$ i.e. $\bar \omega^+ = \lambda_{S}$  
(cf. eqs. (\ref{power}) and (\ref{lambda})) is the leading singularity. 
We also have $\lambda_{S}>>\lambda_{NS}$ since $\gamma^+ >> \Delta P_{qq}(0)$ 
($\gamma^+ \approx 1.18$ and $\Delta P_{qq}(0)=4/3$). This means that  
singlet distributions 
and singlet structure functions dominate over the non-singlet 
ones at low $x$.  Since this singularity is also present in the 
functions $\bar R_{S,g}^{\pm}(\omega,\alpha_s)$ it appears in 
$\Delta \bar p_{S,g}(\omega,Q^2)$  for arbitrary value of the scale $Q^2$.   
It should be noticed that the effect of the double $ln^2(1/x)$ resummation can 
be quite strong and, in particular the exponent $\lambda_S$ can easily become 
of the order of unity. This will be seen more explicitely in the quantitative 
implementation of the 
double $ln^2(1/x)$ contributions which we are going to   discuss below.
In the exact leading double $ln^2(1/x)$ approximation the anomalous 
dimensions $\tilde \gamma^{NS}$ and $\tilde \gamma^{\pm}$ 
and the exponents $\lambda_{NS}$ and $\lambda_{S}$ acquire additional contributions 
due to bremsstrahlung diagrams. Their effect on $\lambda_{NS,S}$ was 
estimated in refs. \cite{BARTNS,BARTS} where it was found that the bremmstrahlung 
terms can enhance $\lambda_{NS}$ by about 4\% and reduce  $\lambda_{S}$ 
by about $10 \%$.\\

In order to make the quantitative predictions one has to constrain the 
structure functions by the existing data at large and moderately small values 
of $x$.  For such values of $x$ however the equations (\ref{dlxns}) and 
(\ref{dlxsg}) 
are inaccurate. In this region one should use the conventional 
Altarelli-Parisi equations with complete splitting functions 
$\Delta P_{ij}(z)$ and not restrict oneself to the effect generated only by 
their $z\rightarrow 0$ part.    
Following refs. \cite{BBJK,BZIAJA} we do therefore extend  equations 
(\ref{dlxns},\ref{dlxsg}) and 
add to their right hand side the contributions coming from the 
remaining parts of the splitting functions $\Delta P_{ij}(z)$.  
We also allow the coupling $\alpha_s$ to run setting $k^2$ as the relevant 
scale. In this way we obtain unified system of equations which contain 
both the complete LO Altarelli-Parisi evolution and the double logarithmic 
$ln^2(1/x)$ effects at low $x$.  
The corresponding system of equations reads:
$$ 
f_{k}(x^{\prime},k^2)=f^{(0)}_{k}(x^{\prime},k^2) + 
{\alpha_s(k^2)\over 2 \pi}{4\over 3}   
\int_{x^{\prime}}^1 {dz\over z} 
\int_{k_0^2}^{k^2/z}
{dk^{\prime 2}\over k^{\prime 2}} 
f_{k}\left({x^{\prime}\over z},k^{\prime 2}\right)+
$$
$$
{\alpha_s(k^2)\over 2\pi}\int_{k_0^2}^{k^2}{dk^{\prime 2}\over
k^{\prime 2}}{4\over 3}\int _{x^{\prime}}^1
{dz\over z} {(z+z^2)f_k({x^{\prime}\over z},k^{\prime 2})-
2zf_{k}(x^{\prime},k^{\prime 2})\over 1-z}+
$$
\begin{equation}
{\alpha_s(k^2)\over 2\pi}\int_{k_0^2}^{k^2}{dk^{\prime 2}\over
k^{\prime 2}}\left[2 +
{8\over 3} ln(1-x^{\prime})\right]f_{k}(x^{\prime},k^{\prime 2})
\label{unifns}
\end{equation}
($k=u_v, d_v, us$),  

$$ 
f_{S}(x^{\prime},k^2)=f^{(0)}_{S}(x^{\prime},k^2) +{\alpha_s(k^2)\over 2 \pi}    
\int_{x^{\prime}}^1 {dz\over z} 
\int_{k_0^2}^{k^2/z}
{dk^{\prime 2}\over k^{\prime 2}} 
{4\over 3} 
f_{S}\left({x^{\prime}\over z},k^{\prime 2}\right) +
$$ 
$$ 
{\alpha_s(k^2)\over 2 \pi}\int_{k_0^2}^{k^2}{dk^{\prime 2}\over
k^{\prime 2}}{4\over 3}\int _{x^{\prime}}^1
{dz\over z} {(z+z^2)f_S({x^{\prime}\over z},k^{\prime 2})-
2zf_S(x^{\prime},k^{\prime 2})\over 1-z}+
$$
$$
{\alpha_s(k^2)\over 2 \pi}\int_{k_0^2}^{k^2}{dk^{\prime 2}\over
k^{\prime 2}}\left[ 2 +
{8\over 3} ln(1-x^{\prime})\right]
f_{S}(x^{\prime},k^{\prime 2}) + 
$$
$$ 
{\alpha_s(k^2)\over 2 \pi}N_F\left[-\int_{x^{\prime}}^1 {dz\over z} 
\int_{k_0^2}^{k^2/z}
{dk^{\prime 2}\over k^{\prime 2}}f_{g}
\left({x^{\prime}\over z},k^{\prime 2}\right) + \int_{k_0^2}^{k^2}{dk^{\prime 2}\over
k^{\prime 2}}\int _{x^{\prime}}^1
{dz\over z} 2z f_g({x^{\prime}\over z},k^{\prime 2})\right] 
$$
 
$$ 
f_{g}(x^{\prime},k^2)=f^{(0)}_{g}(x^{\prime},k^2) + {\alpha_s(k^2)\over 2 \pi}    
\int_{x^{\prime}}^1 {dz\over z} 
\int_{k_0^2}^{k^2/z}
{dk^{\prime 2}\over k^{\prime 2}} 
{8\over 3}
f_{S}\left({x^{\prime}\over z},k^{\prime 2}\right) + 
$$
$$
{\alpha_s(k^2)\over 2 \pi}\left[     
\int_{k_0^2}^{k^2}
{dk^{\prime 2}\over k^{\prime 2}}\int_{x^{\prime}}^1 {dz\over z}  
(-{4\over 3})zf_{S}
\left({x^{\prime}\over z},k^{\prime 2}\right) + 
12   
\int_{x^{\prime}}^1 {dz\over z} 
\int_{k_0^2}^{k^2/z}
{dk^{\prime 2}\over k^{\prime 2}} 
 f_{g}\left({x^{\prime}\over z},k^{\prime 2}\right)\right] +
$$ 
$$
{\alpha_s(k^2)\over 2 \pi}    
\int_{k_0^2}^{k^2}
{dk^{\prime 2}\over k^{\prime 2}} \int_{x^{\prime}}^1 {dz\over z} 
6z\left[{f_{g}
\left({x^{\prime}\over z},k^{\prime 2}\right)- f_{g}
(x^{\prime},k^{\prime 2})\over 1-z} -2f_{g}
\left({x^{\prime}\over z},k^{\prime 2}\right)\right]+
$$
\begin{equation}
{\alpha_s(k^2)\over 2 \pi}    
\int_{k_0^2}^{k^2}
{dk^{\prime 2}\over k^{\prime 2}}\left[ {11\over 2} -{N_F\over 3}
 + 6 ln(1-x^{\prime})\right]f_{g}
(x^{\prime},k^{\prime 2}). 
\label{unifsg}
\end{equation}

The inhomogeneous terms $f_i^{(0)}(x^{\prime},k^2)$ are expressed in terms of the input (integrated) 
parton distributions and are the same as in the case of the LO Altarelli 
Parisi evolution \cite{BBJK}: 
$$
f_k^{(0)}(x^{\prime},k^2)={\alpha_s(k^2)\over 2 \pi}{4\over 3}
\int _{x^{\prime}}^1
{dz\over z} {(1+z^2)\Delta p_k^{(0)}({x^{\prime}\over z})-
2z\Delta p_k^{(0)}(x^{\prime})\over 1-z}+
$$
\begin{equation}
{\alpha_s(k^2)\over 2 \pi}\left[2 +{8\over 3} 
ln(1-x^{\prime})\right]\Delta p_k^{(0)}(x^{\prime}) 
\label{fns0}
\end{equation}
($k=u_v,d_v,us$), 

$$
f_S^{(0)}(x^{\prime},k^2) = {\alpha_s(k^2)\over 2 \pi}{4\over 3}
\int _{x^{\prime}}^1
{dz\over z} {(1+z^2)\Delta p_S^{(0)}({x^{\prime}\over z})-
2z\Delta p_S^{(0)}(x^{\prime})\over 1-z}+
$$
$$
{\alpha_s(k^2)\over 2 \pi}\left[(2 +{8\over 3} ln(1-x^{\prime}))
\Delta p_S^{(0)}(x^{\prime})+
N_F\int _{x^{\prime}}^1{dz\over z} 
(1-2z)\Delta p_g^{(0)}({x^{\prime}\over z})\right] 
$$

$$
f_g^{(0)}(x^{\prime},k^2) ={\alpha_s(k^2)\over 2 \pi}\left[{4\over 3}
\int _{x^{\prime}}^1{dz\over z}(2-z)\Delta p_S^{(0)}({x^{\prime}\over z}) + ({11\over 2} -{N_F\over 3}
 + 6 ln(1-x^{\prime}))\Delta p_{g}^{(0)}
(x^{\prime})\right]+ 
$$
\begin{equation} 
{\alpha_s(k^2)\over 2 \pi}6    
\int_{x^{\prime}}^1 {dz\over z}\left[ 
{\Delta p_{g}^{(0)}
({x^{\prime}\over z})- z\Delta p_{g}^{(0)}
(x^{\prime})\over 1-z} +(1-2z)\Delta p_{g}^{(0)}
({x^{\prime}\over z})\right].
\label{fsg0}
\end{equation}
Equations (\ref{unifns},\ref{unifsg}) together with (\ref{fns0},\ref{fsg0}) and 
(\ref{dpi}) reduce to the LO Altarelli-Parisi evolution equations with the 
starting (integrated) distributions $\Delta p_i^{0}(x)$  
after we set the upper 
integration limit over $dk^{\prime 2}$ equal to $k^2$ in all terms in 
equations (\ref{unifns},\ref{unifsg}) and if we set $Q^2$ in place of $W^2$ as 
the upper integration limit in the integral in eq. (\ref{dpi}).  

%FIGURE 2
\begin{figure}[htb]
   \vspace*{-1cm}
    \centerline{
     \psfig{figure=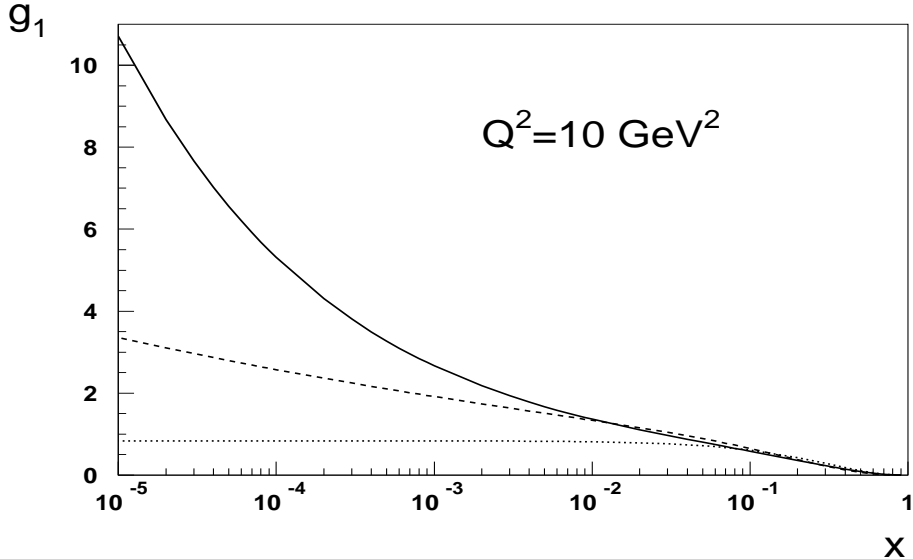,height=12cm,width=14cm}
               }
    \vspace*{-0.5cm}
     \caption{Non-singlet part of the proton spin structure function $g_1(x,Q^2)$
as a function of $x$ for $Q^2=$10 GeV$^2$. Continuous line
corresponds to the  calculations which contain the leading 
$ln^2(1/x)$ resummation, 
, broken line is a  leading order Altarelli--Parisi prediction,
 and a dotted one shows the non-perturbative part t $g_1^{(NS 0)} = 
  g_A/6(1-x)^3$, where $g_A$ denotes the axial vector coupling. The Figure is 
  taken 
  from ref. \cite{BBJK}.}
\label{fig.2}
\end{figure} 
Equations (\ref{unifns},\ref{unifsg}) were solved in refs. \cite{BBJK,BZIAJA} 
 assuming the following simple parametrisation of the 
input distributions: 

\begin{equation}
\Delta p_i^{(0)}(x)=N_i (1-x)^{\eta_i}
\label{dpi0}
\end{equation} 
with $\eta_{u_v}=\eta_{d_v}=3,$  $\eta_{\bar u} = \eta_{\bar s} = 7 $ 
and $\eta_g=5$.  The normalisation constants $N_i$ were determined 
 by imposing the Bjorken sum-rule for $\Delta u_v^{(0)}-\Delta d_v^{(0)}$ 
 and requiring that the first moments of 
all other distributions are the same as those determined from  the recent 
QCD analysis \cite{STRATMAN}. All distributions $\Delta p_i^{(0)}(x)$ 
behave as $x^0$ in the limit $x\rightarrow 0$ that corresponds to the implicit 
assumption that the Regge poles which correspond to axial vector mesons, which 
should control the small $x$ behaviour of $g_1$ \cite{IOFFE,KARL} have their intercept equal 
to $0$.  It was checked that the parametrisation (\ref{dpi0}) combined with 
equations 
(\ref{dpi},\ref{gp1},\ref{unifns},\ref{unifsg}) gives  
reasonable description of the recent SMC data on $g_1^{NS}(x,Q^2)$ and on  
$g_1^p(x,Q^2)$ \cite{SMC}.
%%%%%%%%%%%%FIGURE 3
\begin{figure}[htb]
   \vspace*{-1cm}
    \centerline{
     \psfig{figure=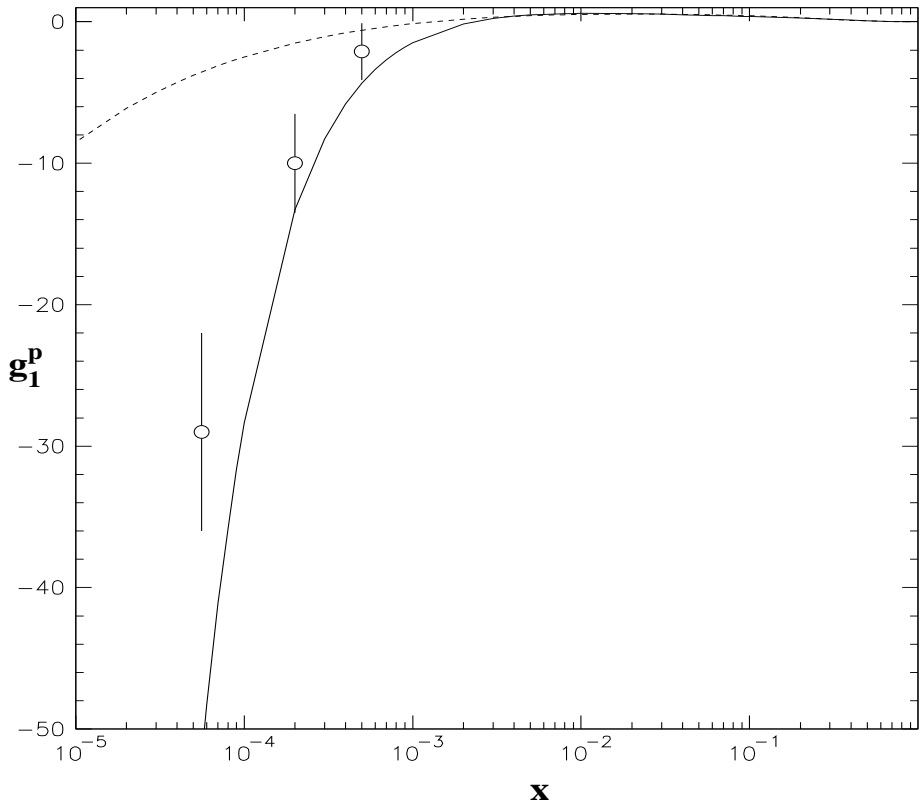,height=12cm,width=14cm}
               }
    \vspace*{-0.5cm}
     \caption{ The structure function $g_1^p(x,Q^2)$ for $Q^2=10 GeV^2$ plotted as 
the function 
of $x$.  Solid line represents this structure function with the double 
$ln^2(1/x)$ terms included and the dashed line corresponds to $g_1^p$ obtained  
from the LO Altarelli-Parisi equations  The "experimental" points are based on 
the NLO QCD predictions with the statistical errors expected at HERA 
\cite{ALBERT}. The Figure is taken from ref. \cite{BZIAJA}.}
\label{fig.3}
\end{figure} 
%%%%%%%%%%%
In Fig.2 we show the nonsinglet part of $g_1(x,Q^2)$ 
for $Q^2= 10 GeV^2$ in the small $x$ 
region \cite{BBJK}. We show predictions based on   equations 
(\ref{unifns},\ref{dpi}) and confront them with the expectations 
which follow 
from solving the LO Altarelli-Parisi evolution equations with the input 
distributions at $Q_0^2=1 GeV^2$ given by equation(s) (\ref{dpi0}). We also 
show the nonperturbative part of the non-singlet distribution 
$g_1^{NS(0)}(x)=g_A/6(1-x)^3$, where $g_A$ is the axial vector coupling. 
In Fig. 3 we show $g_1^p(x,Q^2)$ for $Q^2=10$ where we again confront 
predictions based on equations (\ref{unifns},\ref{unifsg},\ref{dpi}) 
with those based on the LO Altarelli-Parisi evolution equations. 
 We also show in this Figure the "experimental" points which were obtained 
from the extrapolations based on the NLO QCD analysis together with estimated 
statistical errors of possible polarised HERA measurements \cite{ALBERT}.  
We see that the structure function $g_1^p(x,Q^2)$ which contains effects 
of the double $ln^2(1/x)$ resummation begins to differ from that calculated within 
the LO Altarelli Parisi equations already for $x \sim 10^{-3}$.  
It is however comparable to the structure function obtained from the 
NLO analysis for $x>10^{-4}$ which is indicated by the "experimental" points. 
This is presumably partially an artifact of the difference in the input distributions but 
it also reflects the fact that the NLO approximation contains the 
first two terms of the double $ln^2(1/x)$ resummation in the corresponding 
splitting and coefficient functions. It can also be seen from Fig.3 that  the 
(complete) double $ln^2(1/x)$ 
resummation generates the structure function which is significantly steeper than 
that obtained from the  NLO QCD analysis and the difference between those two extrapolations 
becomes significant for $x < 10^{-4}$.\\
%FIGURE 4
\begin{figure}[htb]
   \vspace*{-1cm}
    \centerline{
     \psfig{figure=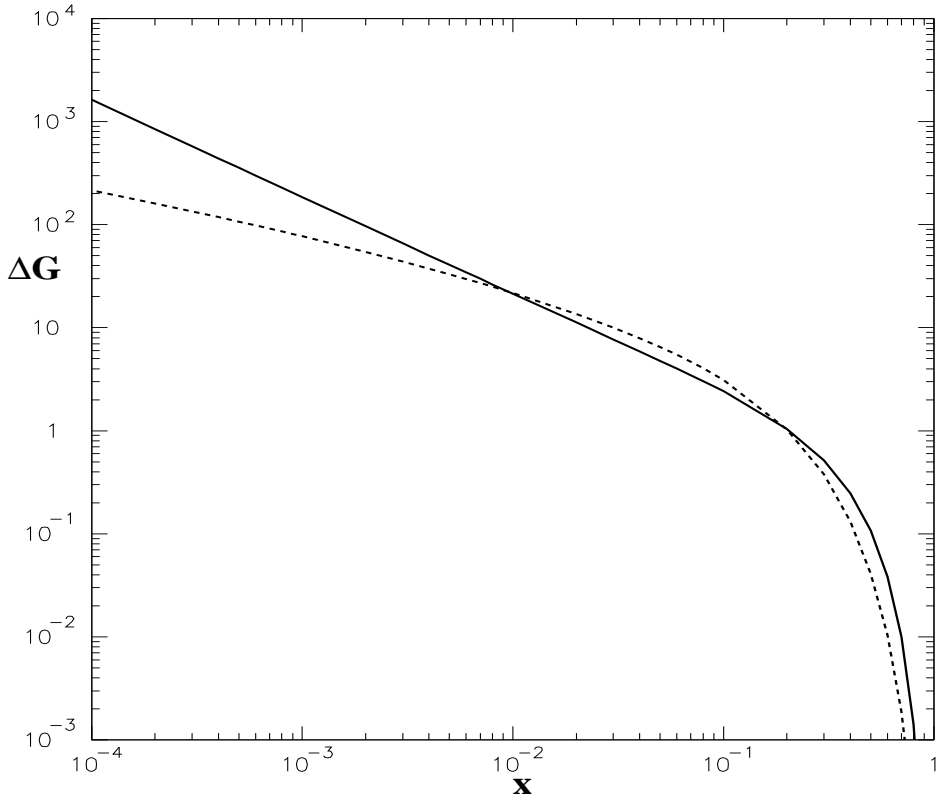,height=12cm,width=14cm}
               }
    \vspace*{-0.5cm}
     \caption{The spin dependent gluon distribution $\Delta G(x,Q^2)$  
for $Q^2=10 GeV^2$ plotted as the function 
of $x$.  Solid line represents $\Delta G(x,Q^2)$   with the double 
$ln^2(1/x)$ terms included and the dashed line corresponds to 
the $\Delta G(x,Q^2)$ 
obtained from the LO Altarelli-Parisi equations.  The Figure is taken from 
ref. \cite{BZIAJA}. }
\end{figure} 
In Fig. 4 we show the spin dependent gluon distribution which 
contains effects of the double $ln^2(1/x)$ resummation 
and confront it with that which was obtained 
from the LO Altarelli-Parisi equations.  It can be seen that the former  
 exhibits characteristic 
$x^{-\lambda_S}$ behaviour with $\lambda_S \sim 1$.  Similar behaviour is also 
exhibited by the structure function $g_1^p(x,Q^2)$ itself.\\

To sum up we have presented theoretical expectations for the low $x$ behaviour of 
the spin dependent structure function $g_1(x,Q^2)$ which follows from the 
resummation of the double $ln^2(1/x)$ terms. 
We have also presented results of the analysis of the "unified" equations 
which contain the LO Altarelli Parisi evolution and the double $ln^2(1/x)$ effects 
at low $x$.  As the first approximation we considered  those double 
$ln^2(1/x)$ effects which are generated by ladder diagrams.  
The double logarithmic effects were found to be very important and they should 
 in principle be visible in possible HERA measurements (cf. Fig. 3). \\
                                                                                                                                                                                                                                                                                                                                                                                                                                                                                                                                        
\section*{Acknowledgments}
I would like to congratulate Marek Je\.zabek  for organizing an excellent 
meeting.    
I  thank Barbara Bade\l{}ek and Beata Ziaja for the most enjoyable 
research collaboration on some of the problems presented in this talk. I thank 
them as well as  
Albert De Roeck and Marco Stratmann for useful 
and illuminating discussions. 
This research has been supported in part by the Polish Committee for Scientific 
Resarch grants 2 P03B 184 10 and 2 P03B 89 13. 

\end{document}